\documentclass[12pt,a4paper]{article}
\usepackage{dcolumn}
\usepackage{bm}
\usepackage{amsmath,amssymb}
\usepackage{booktabs,subfigure}
\usepackage{graphicx,psfrag}
\usepackage{epsfig}
\usepackage{color} 
\usepackage{rotating}
\usepackage{slashed}
\usepackage{geometry}
\usepackage{rotating}
\usepackage{subfigure}
\usepackage{caption}
\usepackage{pstricks}
\usepackage{color}
\usepackage{float}
\usepackage{epsfig}
\usepackage{dcolumn} 
\usepackage{bm} 
\usepackage{cite}
\allowdisplaybreaks[3]
\def \pt {p_{T}}
\def  \met {\not\!\!\pt }
\def \Et {E_{T}}
\def  \MET {\not\!\!\Et }

\numberwithin{equation}{section} 
\def\gsim{\lower0.5ex\hbox{$\:\buildrel >\over\sim\:$}}
\def\lsim{\lower0.5ex\hbox{$\:\buildrel <\over\sim\:$}}

\begin{document}
\begin{flushright}
HRI-RECAPP-2014-007
\end{flushright}

\begin{center}
{\bf \large 
Distinguishing Signatures of top-and bottom-type heavy vectorlike quarks at the LHC}

\vskip 0.8cm

{\bf 
               Aarti Girdhar \footnote{\tt{Electronic address: aarti@hri.res.in}}}

 \medskip
 Department of Physics,
 Dr. B. R. National Institute of Technology, Jalandhar, India\\
 and Regional Centre for Accelerator-based Particle Physics,\\
 Harish-Chandra Research Institute, Chatnaag Road \\ 
                 Jhunsi, Allahabad 211019, India \\
                 \bigskip 
{\bf             %
Biswarup Mukhopadhyaya \footnote{\tt{Electronic address: biswarup@hri.res.in}}}\\
 \medskip
 
Regional Centre for Accelerator-based Particle Physics, Harish-Chandra Research Institute,\\ Chatnaag Road, Jhunsi, Allahabad 211019, India \\
 \bigskip
 {\bf 
                 
 Monalisa Patra \footnote{\tt{Electronic address: monalisa@theory.tifr.res.in}}}\\
\medskip
 Department of Theoretical Physics,\\
 Tata Institute of Fundamental Research, Mumbai 400 005, India \\
\bigskip
\end{center}


\begin{abstract}
\noindent
An $SU(2)$ vectorlike singlet quark with a charge either $+2/3$ $(t')$ or $-1/3$ $(b')$ is 
predicted in many extensions of the Standard Model. The mixing of these quarks with the 
top or bottom lead to Flavor Changing Yukawa Interactions and Neutral Current.
The decay modes of the heavier mass eigenstates are therefore different from the 
Standard Model type chiral quarks. The Large Hadron Collider (LHC) will provide an ideal
environment to look for the signals of these exotic quarks. Considering all decays, 
including those involving $Z$- and Yukawa interactions, we show how one can 
distinguish between $t'$ and $b'$ from ratios of event rates with different
lepton multiplicities. The ability to reconstruct the Higgs boson with a mass 
around 125.5 GeV plays an important role in such differentiation.
\end{abstract}

\section{Introduction}
\label{sec:intro}
The Standard Model (SM) of particle physics has enjoyed a remarkable success
in explaining various experimental data. Occasionally some observations have shown 
deviations from the SM expectation,
but have disappeared later with the increase in statistics. Nonetheless, 
the possibility of new physics being indicated by experimental 
data~\cite{hep-ph/0104024,arXiv:0806.2472} has constantly driven physicists towards an inspired quest.
On the theoretical front,
the SM does have some shortcomings as it has
too many free parameters and offers no answer to some of the fundamental questions. 
An example is the issue of naturalness, or the stability of the Higgs boson 
mass against quadratically divergent corrections. In order to address questions such as this, many theoretical scenarios are exploited, with or without direct connections with the questions asked. The currently
running Large Hadron Collider (LHC) at CERN (and the next generation accelerators)
will hopefully provide us with some clues.
\vspace{.2cm}\\
\noindent
An observation that once created some stir was the measurement of 
the forward-backward asymmetry ($A_b^{FB}$) of 
the $b$ quark~\cite{hep-ph/0104024}, by the LEP experiments.
It showed about 2.9$\sigma$ deviation from the
value predicted by the best fit to the precision electroweak 
observables within the SM~\cite{Phys.Lett.B276.247}. In the leptonic sector 
no such discrepancy was observed. These results have motivated many theories, which resolve 
this disagreement through the 
introduction of new quark degrees of freedom or new possible gauge bosons. 
The study of new quarks mixing with the chiral fermions of the SM has 
been an important area of investigation for quite some time now. 
Many theories beyond the SM (BSM) naturally predict the existence
of such vectorlike quarks.
They have been studied in the
context of the superstring-inspired $E_6$ models~\cite{Phys.Rev.D33.1912}, 
Little Higgs theories ~\cite{hep-ph/0206020,hep-ph/0206021} and 
also extra dimensional models~\cite{hep-ph/0004072}. Many of these scenarios also predict vectorlike leptons. Our focus, however, will be on vectorlike quarks, since they can be expected to be produced at the LHC through strong interactions.
Depending on the 
context, these vectorlike quarks can exist as triplets, doublets 
or singlets under $SU(2)_L$ gauge group and can have different hypercharges 
under $U(1)_Y$. On the other hand the presence of new chiral fermions,
like the one predicted by fourth generation theories, with couplings
similar to the SM ones, are disfavoured by 
data~\cite{arXiv:1209.1062} which allow a narrow mass window for their
survival. The limits on these quarks from various experiments are discussed
later.
\vspace{.2cm}\\
\noindent
In a general context, vectorlike quarks can be both top-like (charge $+{2/3}$) and 
bottom-like (charge $-{1/3}$) vector singlet. 
We consider vectorlike isosinglet quarks in both the sectors as possibilities, taking one 
at a time in addition to the three generations of SM chiral fermions. 
The extension of the SM through 
the inclusion of a weak isospin singlet fermion
leads to mixing between the singlet fermions and the SM doublets and hence 
different phenomenological consequences from what are predicted by the SM.
The aim of our work is to distinguish between the
singlets top- and bottom-type ($t'$ and $b'$ ) from their decays. In particular we wish to utilise the 
fact that the decay into Higgs is possible, and that the mass of the Higgs is known to us now. The dominant
decays of $t'$ and $b'$ are expected to be
\begin{enumerate}
 \item $t'\rightarrow W^+b,~~~t'\rightarrow Ht,~~~t'\rightarrow Zt$.
 \item $b'\rightarrow W^-t,~~~b'\rightarrow Hb,~~~b'\rightarrow Zb$.
\end{enumerate}
 With the discovery of a Higgs like boson by the LHC experiments
\cite{arXiv:1207.7214, arXiv:1207.7235}, the decay mode
$t'/b' \rightarrow H t/b$ is a channel of interest. We make use of 
this fact and tag five $b$'s along with the requirement of two 
$b$ pairs giving invariant mass peaks at $m_H$ for both the isosinglets. 
We find that we can distinguish between the signals present on account 
of $t'$ and $b'$. 
\vspace{.2cm}\\
\noindent
The collider phenomenology of
these vectorlike isosinglets has been considered extensively
in the literature(for most recent
ones~\cite{Phys.Lett.B186.147} -\cite{arXiv:1207.5607}).
In the earlier works, events are mainly selected
with a final state composed of $W$ or $Z$ bosons and jets consistent
with the decay of the heavy quarks. Once a signal is obtained,
it becomes necessary to pinpoint the new physics scenario which leads to it. 
Since both $t'$ and $b'$ mimic the same signal through these channels, we are
addressing this question by our analysis through the Higgs decay channel.
\vspace{.2cm}\\
\noindent
The outline of the paper is as follows. In section~\ref{pheno} we
discuss the couplings of the $t'$ and $b'$ separately
to the SM fields, through the effective Lagrangian in a
model -independent way. In section~\ref{signal} we  discuss the 
signal and the background
along with the  methodology adopted for the analysis of the signal.
In section~\ref{results} the results of our numerical analysis based on 
Monte Carlo simulations is presented. We summarise and conclude in 
section~\ref{discussion}.
\section{Phenomenology of $t'$ and $b'$}\label{pheno}
Strong processes can produce both $t'$ and $b'$-type quarks at the LHC with identical rates, through gluon fusion or quark-antiquark annihilation. Such pair-production, whose rates are independent of the degree of singlet-doublet mixing, are the modes relevant for our study. Though single-production is also possible, perhaps with less phase space suppression, it is\\
(a)~driven by electroweak couplings, and \\
(b)~suppressed by the singlet-doublet mixing angle(s).
\vspace{.2cm}\\
\noindent 
The left and right handed component of the vectorlike quarks have the same 
quantum number under $SU(3) \times SU(2)_L\times U(1)_Y$
unlike the SM ones which are chiral. 
\begin{description}
 \item [I)] $t'_L, t'_R = \left(3, 1, 4/3\right)$ with electric charge  $+2/3$, and mixes with $t$.
 \item [II)] $b'_L, b'_R = \left(3, 1, -2/3\right)$ with electric charge  $-1/3$, and mixes with $b$.
\end{description}
\subsection{Mixing and Coupling of $t'$ and $b'$}\label{Lag}
With the addition of the isosinglet quarks to the SM content,  we assume mixing to take place mainly with
the third generation of the quarks. We show below the general scheme of mixing in the down sector; 
the pattern is similar in the up-sector as well. The weak eigenstates are denoted by $(d_w, s_w, b_w, b'_w)$ and
they are related to the mass eigenstate
by~\cite{Phys.Rev.Lett.66.285,Phys.Lett.B266.112}

\begin{eqnarray}\label{mix_mat1}
 \left(\begin{array}{c} d_w\\ s_w\\ b_w\\ b'_w\end{array}\right)
 &=&U\left(\begin{array}{c} d\\ s\\ b\\ b'\end{array}\right), \nonumber \\
\end{eqnarray}
where
 \begin{eqnarray}\label{mix_mat2}
U_{4 \times 4} =\left(\begin{array}{c}V_{3\times 4}\\X_{1\times 4}\end{array}\right)
= \left(\begin{array}{cccc} V_{ud} &V_{us} &V_{ub} &V_{ub'}\\
 V_{cd} &V_{cs} &V_{cb} &V_{cb'}\\
 V_{td} &V_{ts} &V_{tb} &V_{tb'}\\
 X_{4d} &X_{4s} &X_{4b} &X_{4b'} \end{array}\right).
 \end{eqnarray}

\begin{figure}[H]
\begin{minipage}[b]{0.45\linewidth}
\vspace*{0.45cm}
\centering
\includegraphics[width=6.5cm, height=5cm]{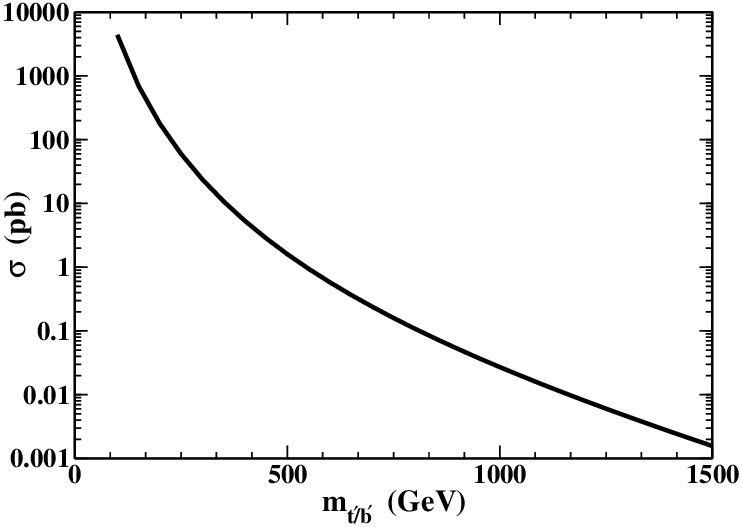}
\caption{The pair production cross section of $t'$ and $b'$ as a function 
of mass for mixing angle $\theta$ = 5 in a 14 TeV LHC.}
\label{fig:csec}
\end{minipage}
\hspace{0.7cm}
\begin{minipage}[b]{0.45\linewidth}
\centering
\includegraphics[width=6.5cm, height=5cm]{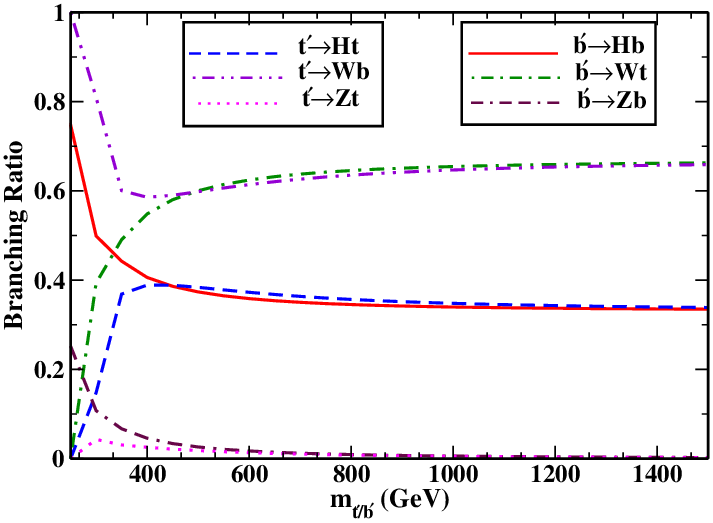}
\caption{The branching ratios for the vectorlike quarks $t'$ and $b'$ as a function
of mass for a fixed mixing angle $\theta$ = 5.}
\label{fig:br}
\end{minipage}
\end{figure}

\noindent The unprimed fields denote the basis, where, the mass matrix of the
up-type quarks are diagonalized. 
The submatrix $V_{3\times 4}$ consisting of the
first three rows of the $U_{4\times4}$ matrix is the charged current matrix 
analogous to the SM CKM matrix and is not unitary. The addition of the fourth row restores the unitarity of U.
The charged current interaction in the mass basis is now given by:
\begin{equation}\label{cc}
 \mathcal{L}_{CC}=\frac{e}{2\sqrt{2}\sin\theta_W}\left[\bar{u}^i_L\gamma^\mu V_{ij} d^j_L\right]W^+_\mu + h.c.
\end{equation}
where $e$ is the electromagnetic
coupling constant, $\theta_W$ is the weak mixing angle and 
$V_{ij}$ is the relevant $3\times 4$ submatrix of U. The indices $i, j$ run over the
quark generations ($i$ = 1-3, $j$ = 1-4), and $u$ = ($u$, $c$, $t$), 
$d$ = ($d$, $s$, $b$, $b'$).
As a consequence of mixing between fields with different weak isospin ($T_3$), 
Flavor Changing Neutral Current (FCNC) processes
appear at the tree level, something that is absent in the SM framework. We 
therefore get
$b'bZ$ and $b'bH$ interactions. In addition, the SU(2) singlet 
field $b'$ in the flavor terms can have a gauge invariant `bare' mass term, 
contrary to $d$, $s$ and $b$. As a result the mass and Yukawa coupling matrices 
cannot be simultaneously diagonalised, and the physical states can have flavor 
changing Yukawa interactions. 
The neutral current interaction in the mass basis is given by:
\begin{equation}\label{nc}
 \mathcal{L}_{NC}=\frac{e}{2 \sin 2\theta_W}\left[\bar{u}^{k}_L\gamma^\mu u_{kL}-
 \bar{d}^i_L \gamma^\mu (V^\dagger V)_{ij}d^j_L-2\sin^2\theta_W J^\mu_{em}\right]Z_\mu
\end{equation}
The index $k$ runs from 1 to 3, whereas $i$ and $j$ runs from 1 to 4.
The electromagnetic current $J^\mu_{em}$ is diagonal in the mass basis and is
\begin{equation}\label{em}
 J^\mu_{em}=\frac{2}{3}\bar{u}^{k}\gamma^\mu u_k-\frac{1}{3}\bar{d}^i\gamma^\mu d_i
\end{equation}
The FCNC coupling as seen from Eq.~(\ref{nc}) is controlled by $V^\dagger V$,
which is a $4\times4$ matrix. Since the mixing matrix $V_{3\times4}$ is embedded 
within the unitary matrix, therefore from Eqs.~(\ref{mix_mat1}),~(\ref{mix_mat2}) we obtain the relation,
\begin{equation}\label{rel}
(V^\dagger V)_{ij}=\delta_{ij}-U^*_{i4}U_{j4}.
\end{equation}
\noindent Now let us come to the full explanation of flavor changing Yukawa 
couplings, already mentioned above. There are gauge invariant
mass terms proportional to $\bar{b}'_Lb'_R$ and $\bar{b}'_Lf'_R$, 
in the Yukawa coupling sector where $f'$ =($d$, $s$, $b$), 
due to non chiral nature of vector quarks.
As these terms do not arise from the Yukawa coupling, therefore, 
the Yukawa matrix cannot be simultaneously diagonalized with the mass
matrix  through a biunitary transformation, consequently giving rise to non 
diagonal Yukawa coupling among the physical quarks. The relations go analogously
for the top sector. The explicit forms of the Yukawa couplings along with the 
coupling of the gauge bosons to the vector quarks and SM quarks 
due to mixing are given as
\begin{eqnarray}\label{coup}
 &&\mathcal{L}_{f\bar{f}H}=\frac{e}{2m_W \sin\theta_W}m_fH\bar{f}(1-|U_{43}|^2)f \nonumber \\
 &&\mathcal{L}_{F\bar{F}H}=\frac{e}{2m_W \sin\theta_W}m_FH\bar{F}(1-|U_{44}|^2)F \nonumber \\
&& \mathcal{L}_{fFH}=\frac{e}{4m_W \sin\theta_W}H\bar{F}(U_{43}^*U_{44}(m_f(1+\gamma_5)+m_F(1-\gamma_5)))f+h.c. \nonumber \\
&&\mathcal{L}_{f\bar{f}Z}=\frac{e\gamma_\mu}{6 \sin 2\theta_W}\bar{f}(3(1-\gamma_5)|U_{33}|^2-4\sin^2\theta^W(|U_{33}|^2+|U_{43}|^2))fZ^\mu \nonumber \\
 &&\mathcal{L}_{F\bar{F}Z}=\frac{e\gamma_\mu}{6 \sin 2\theta_W}\bar{F}(3(1-\gamma_5)|U_{34}|^2-4\sin^2\theta^W(|U_{34}|^2+|U_{44}|^2))FZ^\mu \nonumber \\
 &&\mathcal{L}_{fFZ}=\frac{e\gamma_\mu}{6 \sin 2\theta_W}\bar{f}(3(1-\gamma_5)(U_{33}^*U_{34})
 -4\sin^2\theta_W(U_{33}^*U_{34}+U_{43}^*U_{44}))FZ^\mu+h.c.\nonumber \\
 &&\mathcal{L}_{f'\bar{F}W}=\frac{e}{2\sqrt{2}\sin\theta_W}U_{34}\bar{f'} \gamma_\mu(1-\gamma_5)F W^\mu \nonumber\\
 &&\mathcal{L}_{FFG}=g_s\gamma_\mu \bar{F}FG^\mu
\end{eqnarray}
\noindent where $f$ and $F$ are dominantly $SU(2)_L$ doublet and singlet respectively, 
both in the up and down sectors, and generically 
stand for mass eigenstates. Moreover, $f$ here denotes third generation quarks only.
The CKM matrix elements of the SM involving the light quarks are
directly determined from the experiments and are therefore tightly
constrained. These experimental results also give unitary limits on
the other elements which cannot be directly determined. 
The non-observation of the FCNC decays in the top sector 
by the Tevatron~\cite{Phys.Rev.Lett.80.2525} and the analysis of single top
production in LEP~\cite{Phys.Lett.B521.181} has set bounds on the CKM 
matrix elements involving the top quark at 95\% CL. A detailed analysis on the
allowed mass range and mixing angle $\theta$ in accordance with the precision
electroweak data, flavor physics and oblique parameters is presented 
in~\cite{hep-ph/0210112}. They have
presented the range of the CKM matrix elements allowed for different quark
masses in case of different scenarios. 
For our analysis we have assumed a 
simplified version of the matrix given in Eqs.~(\ref{mix_mat1})  
and~(\ref{mix_mat2}) and describe all the interactions on the addition of
an isosinglet fermion by the following  mixing matrix
\begin{eqnarray}\label{mix_mat3}
 U
=\left(\begin{array}{cccc} V_{ud} &V_{us} &V_{ub} &0\\
 V_{cd} &V_{cs} &V_{cb} &0\\
 V_{td} &V_{ts} &\cos \theta &\sin \theta\\
 0 &0 &- \sin \theta &\cos \theta \end{array}\right).
\end{eqnarray}
\noindent We are considering here the mixing of the vector quark 
with the third generation
only as the  effect of mixing is very small in the lighter generations 
for massive vector quarks. The other elements of the mixing matrix are 
fixed to the SM value. Thus we are essentially considering the isosinglet 
quark in either sector mixing with the third family alone, the mixing angle
being consistent with all existing constraints.
\vspace{.2cm}\\
\noindent
The current phenomenological constraints on vectorlike quarks come 
from direct production bounds at the various colliders and from flavor physics.
There are various direct limits on their masses depending on the decay channel 
analysed. The CDF collaboration has excluded a heavy $t'$ with SM like couplings
at 95\% CL up to 358 GeV~\cite{arXiv:1107.3875} and a heavy $b'$ with SM like 
couplings at 95\% CL up to 372 GeV~\cite{Aaltonen:2011vr}. 
The search mode in the collider experiments is mainly through the pair 
production of these exotic quarks
and further assuming these quarks to only decay through a particular channel.
The bound obtained by CDF on $b'$ was by looking for pair produced heavy quark
with a 100\% branching ratio
to $W$ + SM quarks. This analysis mainly constrains all the models 
predicting this final state.
Recent LHC bounds  from the
ATLAS~\cite{TheATLAScollaboration:2013jha}-\cite{arXiv:1202.3076}
and CMS~\cite{arXiv:1109.4985} data, have set a lower limit on
charge 2/3 and -1/3 exotic quark mass, by the investigation through 
either a particular decay channel or assuming branching ratios
to $W,~Z$ and $H$ decay modes in the context of different models. 
The exclusion of mass of the vectorlike quarks is 
mainly dependent on the strength of their couplings. 
\vspace{.2cm}\\
\noindent
All the possible decay modes of the heavy quarks are 
considered in our analysis. Flavor constraints are also significant for vectorlike quarks. 
The mixing of the new quark with the SM quarks leads to the non-unitarity
of the $3\times 3$ SM CKM matrix and non-unitarity of this 
form is tightly constrained as the unitarity triangle of the SM is being 
measured with absolute precision.
The presence of FCNC in this case contributes to
some processes such as $b\rightarrow s \gamma$, where the quark 
$b$ changes its flavor
by emitting or absorbing $Z$ or Higgs boson. Along with it a 
photon is also emitted.
The FCNC also leads to $b$ meson mixing such as 
$B_d - \bar{B}_d$ and $B_s - \bar{B}_s$ 
mixing~\cite{hep-ph/0210112,arXiv:1207.4440}. 
There being enormous activity in the flavor sector, it is expected that the
experimental data from this sector can be applied to find constraints
on the heavy quarks and their mixing with the SM ones. The constraints
obtained in this case are largely model dependent~\cite{hep-ph/0005133,hep-ph/0102165}
and we do not consider them for our analysis.
The benchmark points that we consider for our calculation are presented 
in Table~\ref{bench}.
\begin{table}[htb]
\begin{center}
\begin{tabular}{||c|c||} \hline
Parameter &Value  \\ \hline \hline
$m_{t'}$ & 350, 500 \\
$m_{b'}$ & 350, 500 \\
$\theta$ & 5, 10, 15 \\ \hline
\end{tabular}
\caption{The various benchmark points used for our analysis, with $\theta$ as 
the mixing angle denoting the $t-t'$ or $b-b'$ mixing.}
\label{bench}
\end{center}
\end{table}
We present in Fig.~\ref{fig:csec} the cross section of the vectorlike quark pair
production at the 14 TeV LHC.  
The main production channels are  gluon-gluon fusion
and $q\bar{q}$ annihilation. The production cross section decreases
with the mass of the vector quarks and is independent of the mixing angle
$\theta$. Moreover the cross section is the same for
the two cases considered here. We next show in Fig.~\ref{fig:br} the 
branching ratios of the various decay
modes of the vector quarks in the two cases plotted as a function of 
their mass. We have kept the Higgs mass fixed at $m_H$ = 125.5 GeV.
These branching ratios are
sensitive to the Higgs mass, and have very weak dependence on $\theta$. 
We have therefore, shown our results for a fixed value of $\theta$.
\section{Signal and Backgrounds}\label{signal}
We consider, as already mentioned, the pair production of both $t' \bar t'$ and 
$b' \bar b'$, taking one at a time via quark-antiquark as well as gluon pair 
annihilation.
\subsection*{Signals}
With the recent discovery of the Higgs with a mass around 125$-$126 GeV, we get
an added edge, given the fact that the Higgs dominantly decays to $b\bar{b}$
in this mass range. With appropriate tagging, it should be possible to 
reconstruct the Higgs from the invariant mass of the $b$-jet pairs. Also, 
for the allowed range of the mixing matrix elements the branching fractions of 
$t'$ or $b'$ to Higgs is substantial for moderate masses.
%
\subsection{Signal for $t'$}
We will be mainly concentrating on the decay mode of $t'$ to Higgs and a top quark.\\
$pp\rightarrow t'\bar{t'}\rightarrow HtH\bar{t}\rightarrow b\bar{b}W^+b b\bar{b}W^-\bar{b}$
\vspace{.2cm}\\
\noindent In this case there can be three possible outcomes
depending on the decay mode of both the $W$'s. It can be either both the $W$'s 
are decaying to leptons or one of them 
is decaying leptonically and the other decaying hadronically, the third 
possibility is the hadronic decay for both the W's. Thus the final states 
arising from $ t'\bar{t'}$ production are
\begin{description}
\item [a)] $6b+2l+\met~~ ({\rm leptonic})$
\item [b)] $6b+1l+\met +2~{\rm jets}~~({\rm semi leptonic})$
\item [c)] $6b+4~{\rm jets}~~({\rm hadronic})$,
\end{description}
\noindent The same final states can also be obtained from other decay modes of 
$t'$. The processes which mimic the $t'\bar{t'}$ decay channel considered for our analysis are 
\begin{description}
\item [a)]  $pp\rightarrow t'\bar{t'}\rightarrow ZtZ\bar{t}\rightarrow b\bar{b}W^+b b\bar{b}W^-\bar{b}$
\item [b)]  $pp\rightarrow t'\bar{t'}\rightarrow ZtH\bar{t}/HtZ\bar{t}\rightarrow b\bar{b}W^+b b\bar{b}W^-\bar{b}$
\end{description}
\noindent The contribution from these decay channels is 
proportional to the branching ratio of $Z \rightarrow b\bar{b}$, which is 
about 15\% and too small. It must be remarked that added to the small branching ratio their 
contribution to the signal gets filtered by the various cuts and 
the calculation of Higgs invariant mass as explained later.
\subsection{Signal for $b'$}
Similar to $t'$, in this case also we will mainly concentrate on the decay mode of
$b'$ to Higgs and bottom quark with the final state consisting of only 6$b$'s. \\
$pp\rightarrow b'\bar{b'}\rightarrow HbH\bar{b}\rightarrow b\bar{b}b\bar{b}b\bar{b}$
\vspace{.2cm}\\
\noindent
There are other modes for $b'$ which can give rise to the final state 
of 6$b$'s. Of course, the contributions from all these processes are
proportional to the respective branching fractions of 
$Z \rightarrow b\bar{b}$. 
\begin{description}
\item [a)] $pp\rightarrow b'\bar{b'}\rightarrow ZbZ\bar{b}\rightarrow b\bar{b}b\bar{b}b\bar{b}$
\item [b)] $pp\rightarrow b'\bar{b'}\rightarrow ZbH\bar{b}/HbZ\bar{b}\rightarrow b\bar{b}b\bar{b}b\bar{b}$
\end{description}
The hadronic decay mode of the $W$'s from $t'$ will give the same final state signature as $b'$, 
provided the jets emitted from the $W$'s are light.
For both $t'$ and $b'$, we look for the following final state signals, mainly with
5 tagged $b$'s reconstructing two Higgs in the mass range, 123$-$128 GeV. 
\begin{description}
\item [Signal 1:] $5b+2l+\met$ 
\item [Signal 2:] $5b+1l+\met$ 
\item [Signal 3:] $5b$ 
\end{description}
The package CalcHEP v2.5.6~\cite{hep-ph/9908288} is used to 
calculate the cross section for the signal process and the respective
branching ratios of $t'$ and $b'$.
%
\subsection*{Backgrounds}
There are many SM processes which can fake the signals listed above. The dominant backgrounds arise from
\begin{enumerate}\label{bg}
\item $pp\rightarrow t\bar{t} H H$
\item $pp\rightarrow t\bar{t}H + 2~ {\rm jets}$ 
\item $pp\rightarrow t\bar{t} + {\rm N~jets}, ~{\rm where} ~0\leq {\rm N} \leq 4$
\item $pp\rightarrow t\bar{t}b\bar{b} + {\rm N~jets}, ~{\rm where} ~0\leq {\rm N} 
\leq 2$ 
\item  $pp\rightarrow W^+W^-HH + {\rm N~jets}, ~{\rm where} ~0\leq {\rm N} \leq 2$
\end{enumerate}
\noindent  We have computed the cross section for all the
background processes except the process 
$pp\rightarrow t\bar{t} H H$, with ALPGEN~\cite{hep-ph/0206293} which takes into
account all the spin correlation 
and finite width effects. The cross section for the production of $t\bar{t}HH$ 
is computed with CalcHEP v2.5.6~\cite{hep-ph/9908288}, and is found to be 
about 0.0005 pb at 14 TeV, for the Higgs mass of 125.5 GeV.
Since it is too small to be a threat to our signal, we are not considering this
process further in our analysis. A similar argument follows for the background 
process $W^+W^-HH$ production in association with jets. The cross section is 
of the order of 10$^{-5}$ pb. Therefore we are also ignoring this process in
the further analysis of the background. The QCD factorisation and 
renormalisation scale ($Q^2$) in ALPGEN  for the different
background processes are presented in Table~\ref{scale}.
\begin{table}[H]
\begin{center}
\begin{tabular}{||c|c||} \hline
Process &$Q^2$  \\ \hline \hline
$pp\rightarrow t\bar{t} H$ & $(2m_t+2m_H)^2$ \\
$pp\rightarrow t\bar{t}$ &$m_t^2$  \\
$pp\rightarrow t\bar{t}b\bar{b}$ &$m_t^2$ \\
$pp\rightarrow W^+W^-HH$ &$m_W+m_H$ \\ \hline
\end{tabular}
\caption{The factorisation and renormalisation scale ($Q^2$) considered
for the different background processes in ALPGEN.}
\label{scale}
\end{center}
\end{table}
\subsection{Event Selection Criteria}\label{cuts}
For the numerical evaluation of both the signal and the background rates, we 
have considered the CTEQ6L parton distribution function with  $m_t$ = 172 GeV,
$m_b$ = 4.8 GeV, $m_H$ = 125.5 GeV
and centre-of-mass energy, $\sqrt{s}$ of 14 TeV. The signal events along with 
their decay branching fractions are generated with the
help of CalcHEP v2.5.6\cite{hep-ph/9908288}. The renormalisation and factorisation scale 
used for the calculation of production cross sections is the default scale used in 
CalCHEP, i.e squared sub process centre-of-mass energy (${m_{ij}}^2=\hat s=({p_i+p_j})^2$). These signal events 
are passed on to PYTHIA-6.4.24 \cite{hep-ph/0603175} for showering and hadronization 
along with the help of CalcHEP-PYTHIA interface program\cite{Belyaev:2000wn}. 
We have taken into account in PYTHIA the initial and final state radiations 
due to QED and QCD, along with the multiple interactions
accounting for pile up. The showering of the SM 
background events is done by passing on the output of ALPGEN~\cite{hep-ph/0206293} 
in the form of unweighted events to PYTHIA.
ALPGEN performs the matching of the jets produced in 
the showering routine to the partons obtained from the matrix element 
calculation using the MLM matching procedure~\cite{Mangano:2006rw}. 
Jet formation is done through FastJet 3.0.2~\cite{arXiv:1111.6097}
using anti-$k_t$ algorithm, with radius parameter $R$ = 0.4. 
The event selection criteria or the cuts applied are the same for both
the signal and the background and are detailed below.
\begin{itemize}
\item Identification of Isolated Leptons (cut 1): 
\vspace*{0.2cm} \\
$1)$ For the lepton trigger, electron candidates are 
required to have $p^e_T > $ 25 GeV 
and $|\eta| <$  2.47. Moreover the electron is vetoed if it lies in the region
$1.37 < |\eta| < 1.52$ between the barrel and endcap electromagnetic 
calorimeters. The muons are required to satisfy $p^\mu_T > $ 25 GeV and 
$|\eta| <$  2.5. 
\vspace*{0.2cm} \\
\noindent
$2)$ Since we are interested in leptons coming from the decay of on-shell 
$W'$s only, they are further tested for being isolated. 
\vspace*{0.1cm}\\
\noindent 
$a)$ The total $E_T$ of stable particles within cone radius 
$\Delta R = \sqrt{(\Delta \eta)^2 + (\Delta \phi)^2}< 0.2$ 
of the lepton should be less than 10 GeV. 
\vspace*{0.1cm}\\
\noindent 
$b)$ In order to make the lepton and jets well separated, we further apply a
lepton jet separation cut, ${\Delta R}_{lj} \geq$ 0.4 on the 
lepton for all the jets formed with $p_T > $ 20 GeV. The jets are formed 
through FastJet~\cite {arXiv:1111.6097},
with $R$ = 0.4 using the anti-$k_t$ jet algorithm.
All the particles other than the leptons with trigger of $p_T > 20$ GeV 
and $|\eta| < $ 2.5 form the input for Fastjet.
The jets trigger for this is $p_T > 20$ GeV.
\vspace*{0.1cm} \\
\noindent 
$c)$ To exclude the contribution of the same flavor leptons that might come from
the decay of $Z$ boson, the  invariant mass $M_{ll}$ of the 
isolated lepton pairs is calculated and the pair 
having mass in the window $|M_Z \pm 10|$ GeV is discarded.
The events chosen after this are listed as those passing cut 1.
For the selection in case of signal with one or two isolated leptons,
after the application of cut 1
all the events with one or two isolated leptons survive.
\item Missing $E_T$ ($\MET$) (cut 2) : \\
For the events with one or two isolated leptons, $\MET$ is calculated by 
computing the vector sum of the visible $p^{tot}_T$ of all particles, where
\begin{eqnarray}\label{mis_et}
 &&\vec{\MET} = -\Sigma\vec{p}^{tot}_T \nonumber \\
 &&p^{tot}_T = \sqrt{(p^{tot}_x)^2+(p^{tot}_y)^2}\nonumber \\
 &&p^{tot}_x = p^{e^\pm}_x+p^{\mu^\pm}_x+p^{jets}_x+p^{unc}_x \nonumber \\
 &&p^{tot}_y = p^{e^\pm}_y+p^{\mu^\pm}_y+p^{jets}_y+p^{unc}_y. 
\end{eqnarray}
In Eq.~(\ref{mis_et}) $p^{unc}_{x,y}$ receives contribution
from the unclustered components,
which consist of the leptons and hadrons in each 
event not passing the primary selection
criteria for trigger but have $p_T>$ 0.5 and $|\eta| < $ 5.0.
A cut of $\MET > $ 40 GeV referred to as cut 2 is applied. 
All the events which survives
cut 1, are subjected to this cut.
\item b tagging (cut 3):\\
The jets  with $E_T > 40$ GeV and $|\eta| < $ 2.5 are selected as trigger for 
the identification of $b$ jets. A jet is tagged as $b$ jet if it 
has a $b$ parton within a cone of $\Delta R < $ 0.4
with the jet axis and a tagging efficiency of 60\% is incorporated. Events 
with five or more $b$'s tagged in this manner are selected and are tabulated as 
events surviving cut 3. 
\item Invariant Mass Reconstruction :\\
For the events with at least 5 tagged $b$'s (surviving after cut 3), 
invariant mass $m_{bb}$ of all possible $b$-jet pairs are computed and those
with $m_{bb}$ in the mass range 123 GeV $\leq m_H \leq$ 128 GeV 
are considered to be coming from the decay of Higgs. We look for the number of 
events which have at least one $b$ pair in the given Higgs
mass range as NH $\ge$ 1 . The number of events with two $b$ pairs within the 
required Higgs mass range are tabulated under NH = 2.  
\end{itemize} 
The cuts mentioned above are mainly motivated to suppress the background
and also to discriminate the signal of the $t'$ and $b'$. 
\subsection{Smearing}
In order to account for detector effects, the momenta of the leptons, jets and 
the unclustered components obtained from the generator are smeared according to 
the following prescription :
\begin{itemize}
\item For electrons and jets: The electrons with pseudorapidity, $|\eta| < $ 2.5 and 
the jets with $|\eta| < $ 5 and $p_T > 20$ GeV are smeared by Gaussian distribution given by
\begin{eqnarray}
\frac{\sigma(X)}{X}=\frac{a}{\sqrt{X}}\oplus b \oplus \frac{c}{X}
\end{eqnarray}
where X = $E_T$.
In case of the electrons
\begin{eqnarray}
(a,b,c) =\left(\begin{array}{cc}(0.030 ~{\rm GeV}^{1/2},~0.005,~0.2 ~{\rm GeV}) &|\eta|< 1.5 \\
                 (0.055 ~{\rm GeV}^{1/2},~0.005,~0.6~{\rm GeV}) &1.5 < |\eta|< 1.5 
                 \end{array} \right),
\end{eqnarray}
whereas for the jets $a = 0.5$ GeV$^{1/2}$, $b$ and $c$ = 0.
\item For muons: Muons with $|\eta| < $ 2.5 are similarly smeared according to
\begin{eqnarray}
\frac{\sigma(p_T)}{p_T}=\left(\begin{array}{cc}a,&p_T<100 ~{\rm GeV}\\
                a+b ~log~\frac{p_T}{100 ~{\rm GeV}},& p_T>100 ~{\rm GeV}
               \end{array}\right)
\end{eqnarray}
with
\begin{eqnarray}
(a,b) =\left(\begin{array}{cc}(0.008,~0.037) &|\eta|< 1.5 \\
                 (0.020,~0.050) &1.5 < |\eta|< 1.5 
                 \end{array}\right)
\end{eqnarray}
\item For unclustered components: The stable particles with $|\eta| < $ 5.0 and $E_T > 0.5$ GeV are smeared as unclustered 
components, the corresponding Gaussian width being 
\begin{eqnarray}
 \sigma(E_T)=\alpha\sqrt{\Sigma_i E_T^{(unc)_i}},
\end{eqnarray}
with $\alpha \approx$ 0.55. In this case the $x$ and $y$ component of $E_T^{unc}$
are smeared independently by the same quantity.
\end{itemize}
\section{Numerical Results}\label{results}
We have briefly discussed in section~\ref{signal} about the 
final state signal along with the various cuts applied for our 
analysis. We present in this section the
actual number of events surviving after each cut for a 
given integrated luminosity of 100 fb$^{-1}$. The different
final states are briefly described below. 
\begin{itemize}
\item First of all, we consider the number of events with the final
state $5b+2l+\met$ (signal 1), i.e. 5 tagged $b$ jets with two $b$ 
jet pair's invariant mass peaking at the Higgs mass 
(123$-$128 GeV) along with 2 isolated leptons, which we call $N_1$.
The number of events, surviving after each cut for this final state, for the 
considered integrated luminosity is presented in Table~\ref{tab:sce1} for both 
the signal and the background processes. 
\vspace{.2cm}\\
\noindent
We can see from the table that with the choice of our cuts, we are able
to discriminate between the signals of the $t'$ and $b'$ vector quarks. From this table we can 
make following observations.
\begin{itemize}
\item At the production level the number of events for both types of signal and one 
of the background, $t\bar{t}$, is of the same order of magnitude. The other two SM 
backgrounds, $t\bar{t}H,~t\bar{t}b\bar{b}$ are smaller but comparable. This continues 
even after the application of cut 1 and cut 2. In case of $b'\bar{b'}$, this is because
the dominant decay mode of $b'$ being $W^{-}t$, the final state will consist of 
4$W$'s and 2$b$'s resulting in large number of events satisfying cut 1 and 2. 
Similarly, the leptons from the $t\bar{t}$ process survives cut 1 and 2 as the tops
produced are highly boosted. It is only after the application of cut 3 
which requires at least 5 tagged $b$'s along with a minimum energy of 40 GeV, the 
background gets washed away.
\item After applying cut 3 the discrimination between the two kinds of signals, $t'\bar{t'}$ and 
$b'\bar{b'}$ starts to show up. Further demand of reconstructing two Higgs from the tagged $b$'s 
makes the distinction between two kinds of signal events quite clear.
\vspace{.1cm}\\
\noindent
We can conclude from this table that in the case of $5b+2l+\met$ final state after all the cuts,
we expect the dominant contribution from top-like vector quark, $t'$. The trend seems to be same 
for both the masses.
\end{itemize}
\begin{table}[htb]
\begin{center}
\begin{tabular}{|c|c|l|c|c|c|c|r|} \hline
&&\multicolumn{6}{|c|}{}\\
 & & \multicolumn{6}{|c|}{Actual Number of Events with $\mathcal{L}$ = 100 fb$^{-1}$} \\[2mm] \cline{3-8}
&&&&&&&\\
$m_{t',b'}$ (GeV) & Process & At prod.& cut 1
&cut 2 & cut 3 & NH $\ge$ 1 & NH=2 \\[2mm] \hline\hline
 &$t'\bar{t'}$ &  1.08 $\times 10^6$ & 3.22 $\times 10^4$ &2.37$\times 10^4$  &179.30 &64.80 &8.64 \\
350 &&&&&&&\\
 &$b'\bar{b'}$ & 1.08 $\times 10^6$ & 3.91$\times 10^4$  & 3.27$\times 10^4$  &23.72 & 5.23 &0.69 \\
&&&&&&&\\ \hline
&$t'\bar{t'}$ & 1.60 $\times 10^5$ &4.92$\times 10^3$  &4.05 $\times 10^3$ &74.69 &18.67 &1.61 \\
500&&&&&&&\\
 &$b'\bar{b'}$ & 1.61 $\times 10^5$ &8.09$\times 10^3$  &7.35 $\times 10^3$   &6.12 &1.41 & 0.1 \\ \hline
    &$t\bar{t}H$ + 2 jets & 1.57 $\times 10^4$ &506.09 &409.08 &0.701 &0 &0 \\
Background  &$t\bar{t}b\bar{b}$ + 2 jets &2.74 $\times 10^4$ &526.75 &426.55 &2.80 &0.60 & 0.06 \\ 
 &$t\bar{t}$ + 4 jets &$1.74 \times 10^6$ & $ 3.46 \times 10^4$ &$2.80 \times 10^4$ &0 & 0 & 0 \\ \hline
\end{tabular}
\end{center}
\caption{Actual number of events in case of the Signal and the Background for a 
$5b+2l+\met$ final state which pass various cuts at the 14 TeV LHC.}
\label{tab:sce1}
\end{table} 
\item We next consider the number of events with at least 5 tagged $b$'s and 1 
isolated lepton i.e. $5b+1l+\met$ (signal 2) in the final state, along with the two $b$
jet pair's invariant mass peaking at the Higgs mass (123$-$128 GeV), 
which we call $N_2$. We present the results for this final state, in Table~\ref{tab:sce2}
for both $t'$ and $b'$ along with the background. The argument in this 
case for the number of events surviving after each cut is similar to
the previous one. The events that survive even after the Higgs invariant mass 
reconstruction in case of backgrounds is mainly due to the combinatorics.
\begin{table}[htb]
\begin{center}
\begin{tabular}{|c|c|l|c|c|c|c|r|} \hline
&&\multicolumn{6}{|c|}{}\\
 & & \multicolumn{6}{|c|}{Actual Number of Events with $\mathcal{L}$ = 100 fb$^{-1}$} \\[2mm] \cline{3-8}
&&&&&&&\\
$m_{t',b'}$ (GeV) & Process & At prod.& cut 1
&cut 2 & cut 3 & NH $\ge$ 1  & NH=2 \\[2mm] \hline\hline
 &$t'\bar{t'}$ &  1.08$\times 10^6$ & 2.92 $\times 10^5$ &2.81$\times 10^5$ &2.56 $\times 10^3$ &743.123 &151.21 \\
350&&&&&&&\\
 &$b'\bar{b'}$ & 1.08$\times 10^6$ & 2.41 $\times 10^5$ & 1.81 $\times 10^5$ &338.7 &74.25 &9.15 \\
&&&&&&&\\ \hline
 &$t'\bar{t'}$ & 1.60 $\times 10^5$ &4.23 $\times 10^4$ &4.26 $\times 10^4$ &928.963 &228.89&33.31 \\
500&&&&&&&\\
 &$b'\bar{b'}$ & 1.61 $\times 10^5$ &4.19 $\times 10^4$ & 3.54 $\times 10^4$ &79.87 &18.2  &3.53 \\ \hline
 &$t\bar{t}H$ + 2 jets & 1.57 $\times 10^4$ &4.20 $\times 10^3$ 
 &3.3 $\times 10^3$ &17.51 &4.55 &0  \\
Background  &$t\bar{t}b\bar{b}$ + 2 jets   &2.74 $\times 10^4$ 
&6.59 $\times 10^3$ &4.75 $\times 10^3$ & 39.24 &8.37 &1.21 \\ 
 &$t\bar{t}$ + 4 jets &1.78 $\times 10^6$ & 4.37 $\times 10^5$ & 3.18 $\times 10^5$ &1.18 $\times 10^2$ 
& 23.76 & 23.76\\ 
\hline
\end{tabular}
\end{center}
\caption{Actual number of events in case of the Signal and the Background for a 
$5b+l+\met$ final state which pass various cuts at the 14 TeV LHC.}
\label{tab:sce2}
\end{table}
\begin{figure}[htb]
\begin{minipage}[b]{0.45\linewidth}
\vspace*{0.45cm}
\centering
\includegraphics[width=6.5cm, height=5cm]{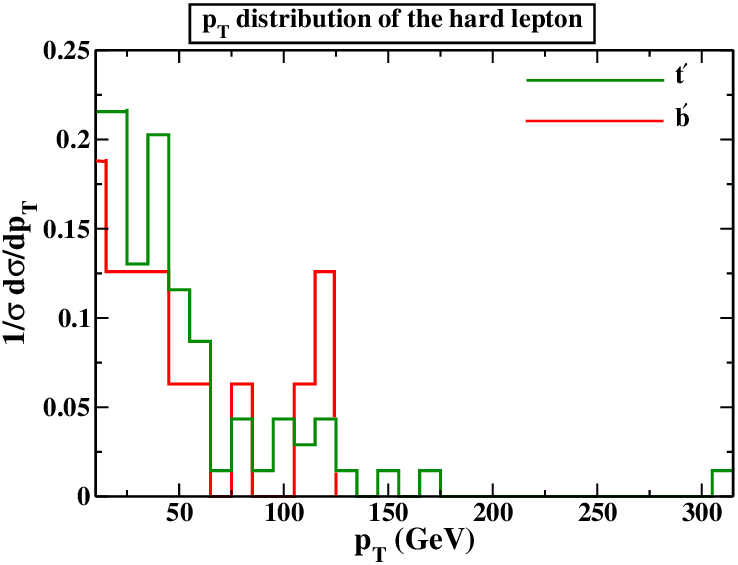}
\caption{The $p_T$ distribution of the hard lepton after imposing all the cuts 
for signal 2 for both $t'$ and $b'$ of mass 350 GeV with $\sqrt{s}$ = 14 TeV.}
\label{fig:pt_lep}
\end{minipage}
\hspace{0.7cm}
\begin{minipage}[b]{0.45\linewidth}
\centering
\includegraphics[width=6.5cm, height=5cm]{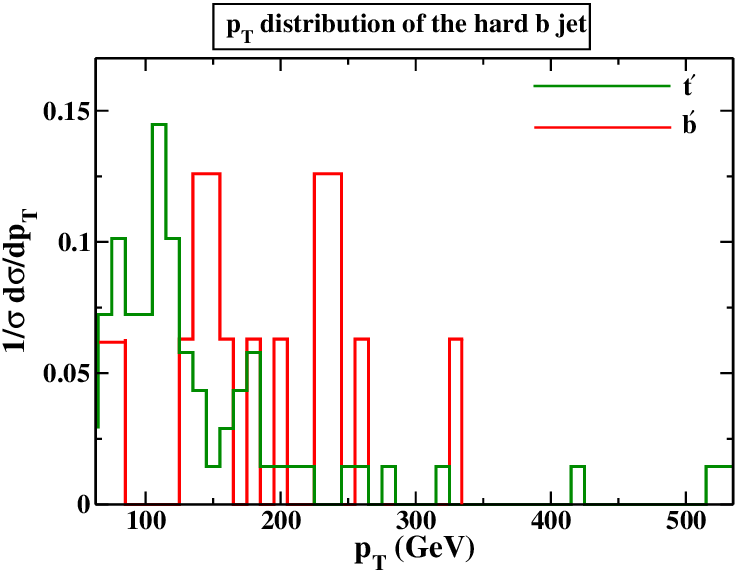}
\caption{The $p_T$ distribution of hardest $b$ jet after imposing all the cuts 
for signal 2 for both $t'$ and $b'$
of mass 350 GeV at $\sqrt{s}$ = 14 TeV.}
\label{fig:pthard}
\end{minipage}
\end{figure}
\begin{figure}[htb]
\begin{minipage}[b]{0.45\linewidth}
\vspace*{0.45cm}
\centering
\includegraphics[width=6.5cm, height=5cm]{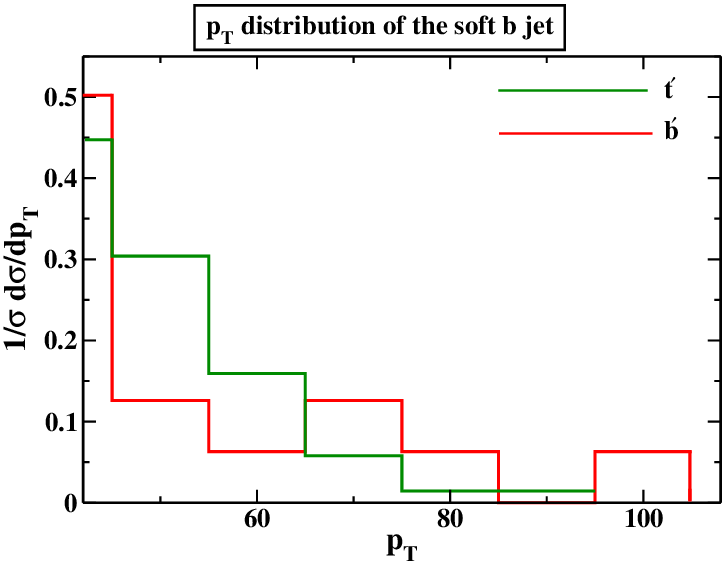}
\caption{The $p_T$ distribution of the soft $b$ jet after imposing all the cuts 
for signal 2 for both $t'$ and $b'$
of mass 350 GeV at $\sqrt{s}$ = 14 TeV.}
\label{fig:ptsoft}
\end{minipage}
\hspace{0.7cm}
\begin{minipage}[b]{0.45\linewidth}
\centering
\includegraphics[width=6.5cm, height=5cm]{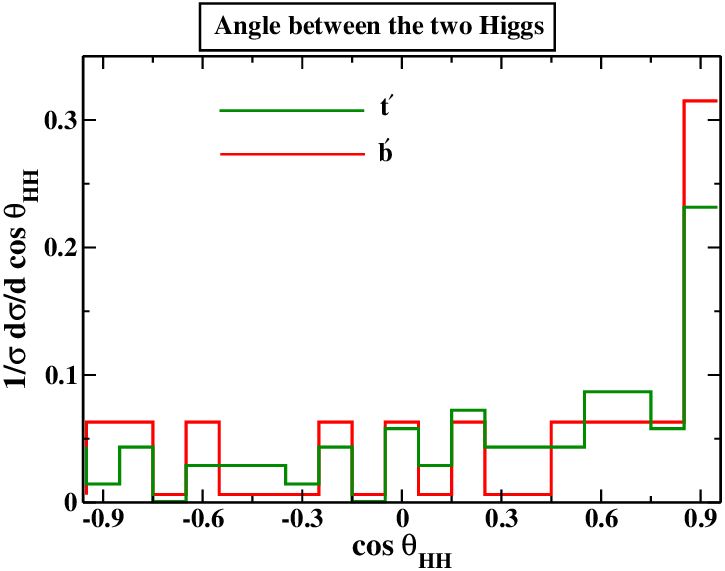}
\caption{The distribution of opening angle between two reconstructed Higgs 
after imposing all the cuts for signal 2 for both $t'$ and $b'$
of mass 350 GeV at $\sqrt{s}$ = 14 TeV.}
\label{fig:anghiggs}
\end{minipage}
\end{figure}
We next show in Figs.~\ref{fig:pt_lep},~\ref{fig:pthard},~\ref{fig:ptsoft}
and~\ref{fig:anghiggs}, the different kinematic distributions for this particular signal
after all the cuts have been imposed.
The $p_T$ distribution of the hard isolated lepton is shown in Fig.~\ref{fig:pt_lep},
whereas Fig.~\ref{fig:pthard} shows the $p_T$ distribution of the hard $b$ jet. 
Similarly the $p_T$ distribution of the soft $b$ jet is shown in Fig.~\ref{fig:ptsoft},
with Fig.~\ref{fig:anghiggs} showing the opening angle between the two reconstructed 
Higgs. The $p_T$ of the isolated lepton mostly varies between 40$-$80 GeV, whereas the 
$p_T$ of the hard $b$ jet varies over a wide range of 80$-$150 GeV. The $p_T$
of the soft $b$ jet is found to be less than 80 GeV with a peak around 50 GeV.
It is seen from the Figures that the distribution pattern of both $t'$ and $b'$
are the same, only with a difference in the statistics. In Figure \ref{fig:anghiggs}, however, a mild
difference is noticed. This is because the lepton final states do not arise in the lowest 
order in $pp \rightarrow b^{\prime}\bar{b}^{\prime} \rightarrow b\bar{b}HH$. The signal is thus the 
consequence of initial/final states radiation and of cases 
when it actually arises from $b^{\prime}\rightarrow tW$, which accidentally mimic two Higgses in an uncorrelated manner.
\item Finally the events with at least 5 tagged $b$ jets and zero leptons (signal 3), 
along with the two $b$ jet pair's invariant mass peaking at the Higgs mass 
(between 123 GeV and 128 GeV), which we call $N_3$, is considered.
The result for this is presented in Table~\ref{tab:sce3} for both $t'$ and $b'$.
While considering the zero lepton final state given in Table~\ref{tab:sce3}, the cut consisting of 
isolated lepton and missing energy i.e. cut 1 and 2 is 
neglected for the obvious reasons. Since both $t'$ and $b'$ favour the hadronic decay mode,
it can be seen from the Table, the number of events surviving is the same 
even after cut 3.
It is only after the Higgs mass reconstruction from the $b$ jets, that both $t'$
and $b'$ show different behaviour. Similar to signal 2, we have shown in 
Figs.~\ref{fig:pthard1},~\ref{fig:pthard2},~\ref{fig:softbb},~\ref{fig:anghiggs2}
the various distributions with all the kinematic cuts imposed. The $p_T$ of the 
hard $b$ jet is peaked around 200 GeV for $b'\bar{b'}$ pair production, whereas
in case of $t'\bar{t'}$, the $p_T$ is peaked around 100 GeV as can be seen from Fig.~\ref{fig:pthard1}.
This is mainly because the $b$'s from $b'\bar{b'}$ are directly produced from the decay of $b'$,
compared to those from $t'\bar{t'}$, where $b$ is produced from the decay of $t'$ to $tH/tZ$ followed 
by the decay of $t$ to $Wb$ or $H/Z$ to $b\bar{b}$. The discrimination is reduced when one moves to
the $p_T$ distribution of the softer jets, Fig.~\ref{fig:pthard2}, with both $t'\bar{t'}$ and $b\bar{b'}$
behaving similarly for the $p_T$ distribution of the soft $b$ jet, Fig.~\ref{fig:softbb}. 
The distribution of the opening angle between the two reconstructed Higgs is shown in
Fig.~\ref{fig:anghiggs2}. The distribution pattern is the same for both $t'$ and $b'$.
It can be seen from the figure that most of the reconstructed Higgs pair have small opening angles,
even though the Higgses reconstructed in the signal are also on account of combinatorics. This
is because these are real signals, as opposed to the events in Figure \ref{fig:anghiggs}, and the boost of the parton center-of-mass frame causes a small opening angle in a large number of cases.
\begin{table}[htb]
\begin{center}
\begin{tabular}{|c|c|l|c|c|r|} \hline
&&\multicolumn{4}{|c|}{}\\
 & & \multicolumn{4}{|c|}{Actual Number of Events with $\mathcal{L}$ = 100 fb$^{-1}$} \\[2mm] \cline{3-6}
&&&&&\\
$m_{t',b'}$ (GeV) & Process & At prod.
 & cut 3 & NH $\ge$ 1 & NH=2 \\[2mm] \hline\hline
 &$t'\bar{t'}$ &  1.08$\times 10^6$  &1.25 $\times 10^4$ &3.71 $\times 10^3$  &704.26 \\
350&&&&&\\
 &$b'\bar{b'}$ & 1.08$\times 10^6$ &4.48 $\times 10^4$ &1.11 $\times 10^4$ &1.19 $\times 10^3$  \\
&&&&&\\ \hline
 &$t'\bar{t'}$ & 1.60 $\times 10^5$ &3.74$\times 10^3$  &9.67 $\times 10^2$ &135.25 \\
500&&&&&\\
 &$b'\bar{b'}$ & 1.61 $\times 10^5$ &6.88 $\times 10^3$  &1.23 $\times 10^3$ &110.2 \\ \hline
 &$t\bar{t}H$ + 2 jets & 1.57 $\times 10^4$  &86.51 &26.62 &3.85 \\
Background  &$t\bar{t}b\bar{b}$ + 2 jets &2.74 $\times 10^4$  &194.58 &42.69 &5.78  \\ 
 &$t\bar{t}$ + 4 jets & $1.78 \times 10^6$ & $7.36 \times 10^2$& 118.78& 0 \\  \hline
\end{tabular}
\end{center}
\caption{Actual number of events surviving after various cuts in case of the Signal and the Background for a 
$5b$ final state with two $b$ pairs giving an invariant mass peak at the 14 TeV LHC.}
\label{tab:sce3}
\end{table} 
\begin{figure}[htb]
\begin{minipage}[b]{0.45\linewidth}
\vspace*{0.45cm}
\centering
\includegraphics[width=6.5cm, height=5cm]{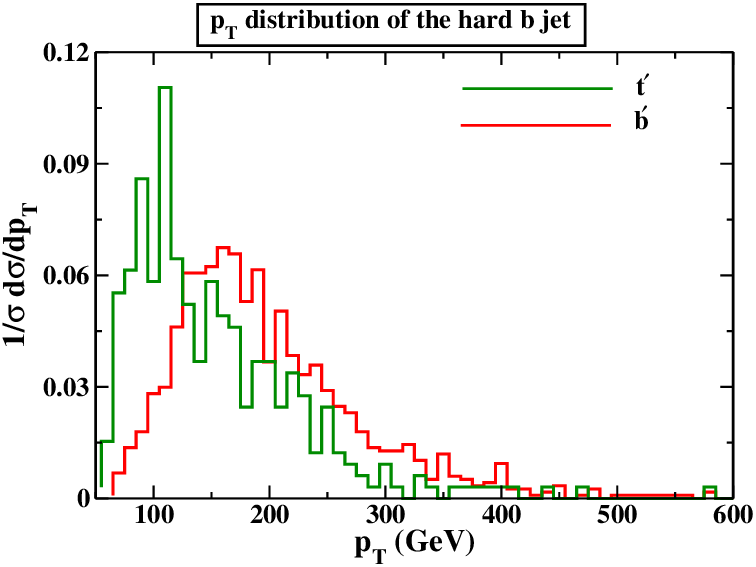}
\caption{The $p_T$ distribution of the hardest $b$ jet after imposing all the cuts 
for signal 3 for both $t'$ and $b'$
of mass 350 GeV at $\sqrt{s}$ = 14 TeV.}
\label{fig:pthard1}
\end{minipage}
\hspace{0.7cm}
\begin{minipage}[b]{0.45\linewidth}
\centering
\includegraphics[width=6.5cm, height=5cm]{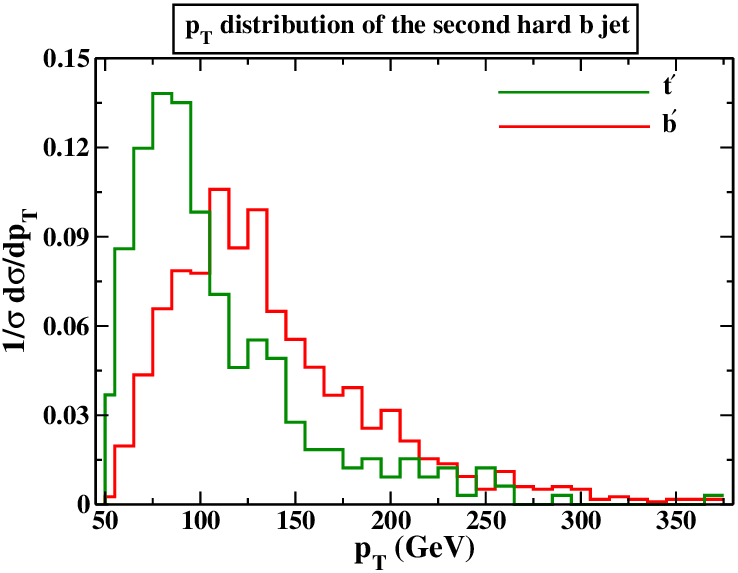}
\caption{The $p_T$ distribution of second hardest $b$ jet after imposing all 
the cuts for signal 3 for both $t'$ and $b'$
of mass 350 GeV at $\sqrt{s}$ = 14 TeV.}
\label{fig:pthard2}
\end{minipage}
\end{figure}
\begin{figure}[htb]
\begin{minipage}[b]{0.45\linewidth}
\vspace*{0.45cm}
\centering
\includegraphics[width=6.5cm, height=5cm]{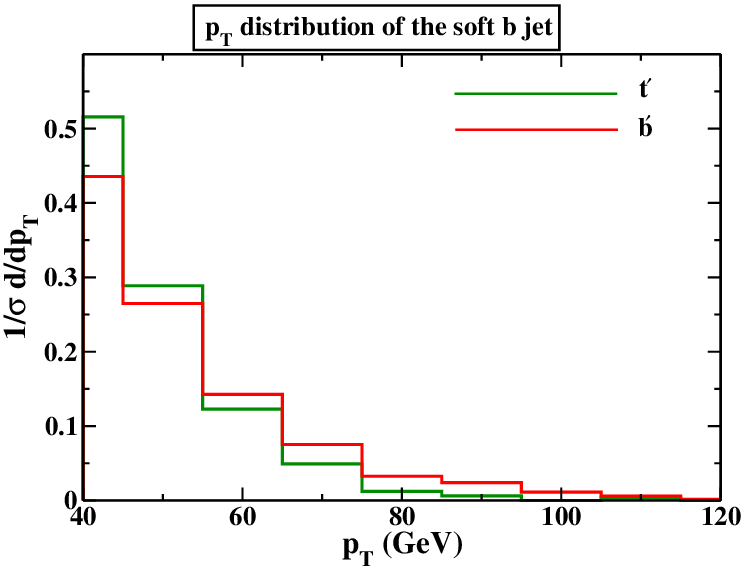}
\caption{The $p_T$ distribution of the soft $b$ jet 
after imposing all the cuts for signal 3 for both $t'$ and $b'$
of mass 350 GeV at $\sqrt{s}$ = 14 TeV.}
\label{fig:softbb}
\end{minipage}
\hspace{0.7cm}
\begin{minipage}[b]{0.45\linewidth}
\centering
\includegraphics[width=6.5cm, height=5cm]{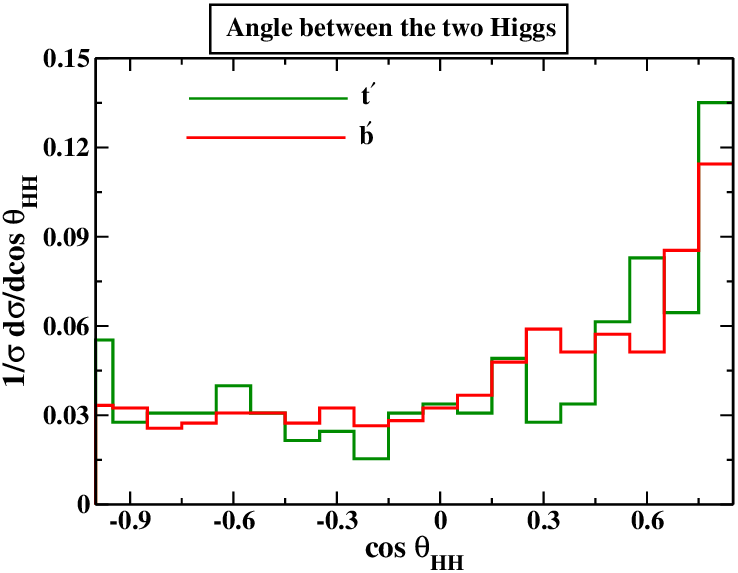}
\caption{The distribution of opening angle between two reconstructed Higgs
after imposing all the cuts for signal 3 for both $t'$ and $b'$
of mass 350 GeV at $\sqrt{s}$ = 14 TeV.}
\label{fig:anghiggs2}
\end{minipage}
\end{figure}
%
%
\end{itemize}
From the above analysis, we find a significant difference in the number of events
that survive after all the cuts, for both the signals. Still we choose to compute the 
ratio of the number of events
surviving after the application of all cuts in case of the different
signals $N_{i}$, where $i=$ 1, 2, 3 as defined before for both $t'$ and $b'$. 
The relevant ratios which we consider are $N_{13}=N_1/N_3$ and 
$N_{23}=N_2/N_3$ for both $t'$ and $b'$. We consider these ratios as
working with the above rates helps us in getting most of systematic uncertainties 
cancelled. The results are presented in Table~\ref{tab:ratio} and it can be seen that they
differ significantly for $t'$ and $b'$. 
\begin{table}[htb]
\begin{center}
\begin{tabular}{|c|c|c|c|} \hline
Mass(GeV)& Isosinglet & $N_{13}$ &$N_{23}$\\
\hline
\hline
& $t'$ &  0.012  &  0.215 \\
350&&&\\
&$b'$ &0.0006 & 0.008\\
\hline
& $t'$ &  0.012 &0.246\\ 
500&&&\\
& $b'$ & 0.0009&0.032 \\ \hline
\end{tabular}
\end{center}
\caption{The ratios, $N_{13}$ and $N_{23}$ for $t'$ and $b'$}
\label{tab:ratio}
\end{table} 
We can make the following observations.
\begin{itemize}
\item The ratio $N_{13}$ differs for $t'$ and $b'$ by two orders of magnitude, for 
both the masses of 350 and 500 GeV. 
\item The ratio $N_{23}$ differs by a factor 
of 100 in case of 350 GeV and by a factor of 10 in the case of 500 GeV. 
\end{itemize}
It turns out that the ratio $N_{13}$ is a better distinguishing observable 
than $N_{23}$ and it continues to be so even when the mass of $t'$, $b'$ increases
while $N_{23}$ seems to be  sensitive to the $t'$, $b'$ mass, and can only be used as
a distinguishing observable for quarks masses unto 700 GeV.
%
%
\section{Summary and Conclusions} \label{discussion}

In this work we have made an attempt to distinguish between top-like and bottom-like
isosinglet quarks which are predicted in several extensions of the SM, at a luminosity 
of 100 fb$^{-1}$ 
and center-of-mass energy of 14 TeV at the LHC. On account of being vectorlike 
they mix with third generation 
chiral quarks which leads to flavor changing Yukawa interactions along with FCNC.
These quarks have the decay modes, $t'\rightarrow Zt,~Ht,~W^{+}b$ and $b'\rightarrow Zb,~Hb,~W^-t$. 
We have in this work tried to address the question of distinguishing 
the signatures of these isosinglet vectorlike quarks once they are discovered.

Choosing in particular the Higgs decay channel out of these possibilities for both $t'$ and $b'$, 
we tried to make a distinction between the two cases. The Higgs decays further to a pair of $b$ quarks. 
We demand that the two Higgs be reconstructed in the mass range 123$-$128 GeV from the tagged $b$'s. 
The recent discovery of Higgs like resonance at 125.5 GeV at the LHC strengthens our analysis. 
We choose three final states with 2, 1, 0 lepton along with five tagged $b$'s which is 
attainable at the LHC as it can efficiently detect leptons and also tag $b$'s.
We find that with a suitable choice of cuts, the SM background is very small for both the signals. 
Our study overall reveals that, empowered by our recent information on the Higgs, 
we can clearly differentiate between $t'$ and $b'$ from ratios of events 
with various lepton multiplicities in the final state along with two reconstructed
Higgs. 
\section{Acknowledgement}
AG is thankful to Department of Science and Technology, New Delhi,
for funding the work done here through the Women in Science scheme
(grant number-(WOS-A/PS-04/2009)). 
BM acknowledges the funding available from the Department of Atomic Energy, 
Government of India, for the Regional Centre for Accelerator based Particle Physics(RECAPP), 
Harish-Chandra Research Institute.
MP would like to extend her thanks to RECAPP for the local hospitality and computational
assistance during the course of this work. 

%
%

\end{document}